\documentclass[12pt]{article}
\usepackage[bottom=2cm, top=2cm, left=2cm, right=2cm]{geometry}
\usepackage{color}

\usepackage{subfigure}
\usepackage{amsmath}
\usepackage[utf8]{inputenc}
\usepackage{standalone}
\usepackage{tikz}
\usepackage{amssymb}
\usepackage{eurosym}
\usepackage{hyperref}
\usepackage{caption}
\definecolor{b}{rgb}{0,0,.4}	%%omega-blau
\definecolor{g}{rgb}{0,.3,0}	%%Tau-grün
\definecolor{n}{rgb}{0,0,0}	%%normal-schwarz
\definecolor{h}{rgb}{0.4,0.2,0.2}	%%hint
\definecolor{v}{rgb}{0.2,0.6,0}

\newcommand{\A}{{\mathbb A}}
\newcommand{\B}{{\mathbb B}}

\newcommand{\bZ}{\boldsymbol Z}

\newcommand{\bbeta}{\boldsymbol \beta}

\DeclareMathOperator*{\argmin}{arg\,min}	% argmin
\DeclareMathOperator{\Var}{Var}
\DeclareMathOperator{\SD}{SD}

\newcommand{\ov}\overline
\newcommand{\what}{\widehat}
\newcommand{\wtilde}{\widetilde}

\definecolor{gray}{rgb}{0.5,0.5,0.5}
\definecolor{red}{rgb}{0.8,0,0}
\definecolor{dred}{rgb}{0.5,0,0}
\definecolor{blue}{rgb}{0,0.1,1}
\definecolor{dblue}{rgb}{0,0.1,0.6}
\definecolor{cyan}{rgb}{0,0.7,.2}
\definecolor{dcyan}{rgb}{0,0.5,.5}

\usetikzlibrary{arrows,positioning,shadows, calc, intersections, fit, automata}
\usetikzlibrary{arrows.meta}
\usetikzlibrary{decorations.pathreplacing}
 	\newcommand{\WM}{\textcolor{white}{-}}
\begin{document}

	\newpage
	
	\begin{center}
		{\large \textbf{The Impact of Renewable Energy Forecasts on Intraday Electricity Prices }}
		
		Sergei Kulakov\footnote{University of Duisburg-Essen, corresponding author. Email: sergei.kulakov@uni-due.de.} \& Florian Ziel\footnote{University of Duisburg-Essen}
	\end{center}
	\begin{abstract}
	{In this paper we study the impact of errors in wind and solar power forecasts on intraday electricity prices. We develop a novel econometric model which is based on day-ahead wholesale auction curves data and errors in wind and solar power forecasts. The model shifts day-ahead supply curves to calculate intraday prices. We apply our model to the German EPEX SPOT SE data. Our model outperforms both linear and non-linear benchmarks. Our study allows us to conclude that errors in renewable energy forecasts exert a non-linear impact on intraday prices. We demonstrate that additional wind and solar power capacities induce non-linear changes in the intraday price volatility. Finally, we comment on economical and policy implications of our findings. }
		
		%(recorded in the Epex in our case)
		
	\end{abstract}

	\section{Introduction}
	
	\subsection{Literature review}
	
	In an effort to curb climate change, contemporary government policies actively promote, support and even force an increased use of clean power. As a result, structural changes to energy system are occurring. The renewable revolution is mainly driven by two technologies: wind and solar. Multiple indicators predict their booming future due to their continuously falling costs, widespread availability and low global warming potential. Yet, energy harnessed by wind turbines or photovoltaic panels is intermittent. {This signifies the importance of load and price forecasting.}

\begin{figure}[h]
		\centering
			\begin{center}
			\begin{tikzpicture}[scale=1.3]
			\node[align=center] at (1,0.5) {};
			\node[align=center] at (1,1) {\small Day-ahead\\[-0.1em]\small auction};
			\node[align=center] at (1,-1) {\small 12:00\\[-0.1em]\small $t-1$};
			\draw [line width=0.25mm] (1,0.25) -- (1,-0.25);
			%\node[align=center] at (3.15,0.85) {Latent\\[-0.1em]period};
			%\draw [thick,decorate,decoration={brace,amplitude=6pt,raise=0pt}] (2.5,0.15) -- (3.75,0.15);
			\draw [line width=0.25mm,] (4,0.25) -- (4,-0.25);
			\node[align=center] at (4,1.28) {\small Start of the\\[-0.1em]\small intraday\\[-0.1em]\small trading};
			\node[align=center] at (4,-1) {\small 15:00\\[-0.1em]\small $t-1$};
			%\node[align=center] at (6.25,0.6) {Infectious period};
			\draw [line width=0.25mm] (9,0.25) -- (9,-0.25);
			\node[align=center] at (9,1.75) {\small Intra-day\\[-0.1em]\small gate closure};
			\draw [line width=0.25mm, ->,  -{Latex[length=2mm]}] (9,1.2) -- (9,0.4);
			\node[align=center] at (9.15,0.5) {};
			\draw [line width=0.25mm,decorate,decoration={brace,amplitude=10pt,raise=0pt}] (4.1,0.15) -- (8.9,0.15);
			\node[align=center] at (6.5,1.28) {\small  Continuous \\[-0.1em] \small intraday\\[-0.1em]\small trading };
			\node[align=center] at (10,0.7) {\small Delivery};
			\node[align=center] at (10,-0.8) {\small$t$};
			\draw [line width=0.25mm] (10,0.25) -- (10,-0.25);
			\draw [line width=0.25mm , ->, -{Latex[length=3mm]}] (0,0) -- (11,0);
			\draw [line width=0.25mm,decorate,decoration={brace,amplitude=6pt,raise=0pt,mirror}] (9.1,-0.15) -- (9.925,-0.15);
			\draw [line width=0.25mm,->,  -{Latex[length=2mm]}] (9.5,-1.2) -- (9.5,-0.5);
			\node[align=center] at (9.5,-1.5) {\small 30 minutes};
			%\node[align=center] at (6.25,-0.85) {Symptomatic\\[-0.3em]period};
			\end{tikzpicture}
		\end{center}
		\caption{{Simplified trading scheme of the German EPEX SPOT SE.}}
		\label{FIG0}
	\end{figure}
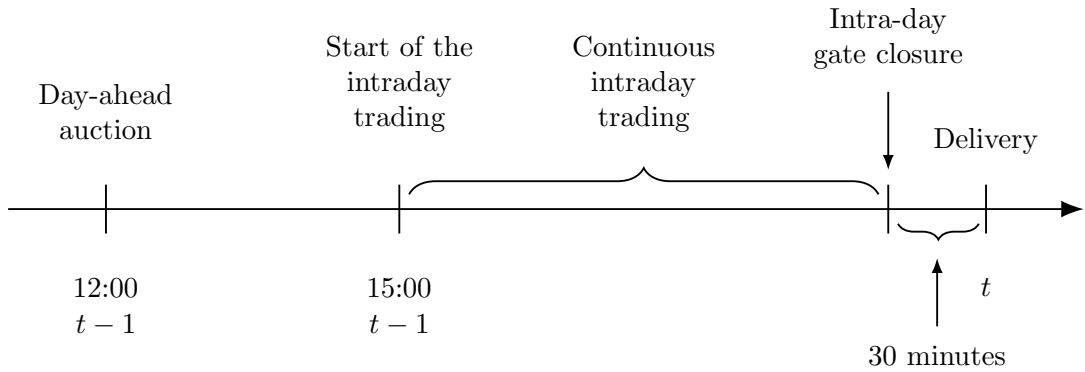

	Variability of the sun and wind energy is critical in the German EPEX SPOT SE. A simplified temporal trading scheme of this energy exchange is depicted in Figure \ref{FIG0}. Two markets are of particular interest to us: day-ahead and continuous intraday. They differ in their temporal proximity to the point $t$ of physical electricity delivery and in their microstructures. The former market is a non-continuous limit order book auction conducted at 12:00 a day prior to $t$. The latter one is a continuous trading market which closes 30 minutes prior to $t$. 
	
	Note that prices in both markets are announced before the physical delivery of electricity occurs. Therefore, the prices are based on wind and solar power supply forecasts. Furthermore, forecasts are usually more precise in the intraday market. Hence, prices in the intraday market are closer to the actual fundamental equilibrium if market participants actively trade in both markets (see e.g. \cite{weber2010adequate} or \cite{pape2017impact}). {As a consequence, the influence of forecast errors on intraday prices drops the closer the trading occurs to the point of actual electricity delivery.  Please ntoe that in this paper we always mean forecast errors in wind and solar power forecasts when we refer to forecast errors. Moreover, note that the impact of a forecast error on an intraday price may depend on the size of the error. Therefore, this paper attempts to prove that the impact of forecast errors on intraday prices is non-linear.}

	The influence of forecast errors on electricity market prices has already been given a thorough attention in the academic literature. \cite{von2009interaction} have explicitly demonstrated the importance of errors in wind forecasts and attempted to measure the impact of these errors on intraday prices. In a more recent study,  \cite{kiesel2017econometric} show that intraday prices are indeed affected by errors in renewable energy forecasts and are even sensible to the signs of forecast errors. \cite{pape2017impact}  shows that errors in wind and solar power forecasts affect not only intraday prices but also controllable electricity producers. \cite{garnier2015balancing} take the perspective of an operator who tries to compensate forecast errors in a continuous intraday market. In doing so, the authors of the paper attempt to determine the optimal timing and volume decisions of an operator. The study of \cite{gurtler2018effect} uses panel data analysis and supports the conclusions drawn by \cite{kiesel2017econometric}. 
	%Khabibrakhmanov, et. al (2016)  signifies the importance of asymmetric effects in errors in solar power on the grounds of the US data.
	The core findings of the latter paper regarding asymmetries are though disputed by \cite{ziel2017modeling} who agrees that forecast errors influence intraday market prices, however, the asymmetric effects are insignificant.  
	
	Non-linearities in the impact of forecast errors have also been analyzed in the academic literature. By investigating the Nord Pool data, \cite{jonsson2010market} show that  electricity prices react on adjustments in wind power predictions more strongly during the day than during the night. Furthermore, the study claims that the price decreases drop when the level of wind penetration rises. \cite{hagemann2013price} relies on the German electricity data and demonstrates that the impact of wind forecasts on electricity prices is more significant from midnight to 8 a.m.  {due to office hours and forecast updates arriving.} {Moreover, many fundamental models which were applied to study the influence of forecast errors (e.g. \cite{goodarzi2019impact} or \cite{zareipour2009economic} besides the above mentioned) are non-linear by nature.}
	
	{Nevertheless, the non-linear impact of errors in wind and solar power forecasts {particularly} on intraday electricity prices has not yet been studied.} The present paper attempts to solve this problem and demonstrates that the impact of forecast errors on intraday prices depends on e.g. the sector of the merit-order curve in which intraday prices are realized. {More importantly, one of the key innovation of the present paper is an auction-curves-based  non-linear econometric model we develop. The core of the model is built around manipulations with empirical supply and demand curves (also known as sale and purchase curves) recorded in a day-ahead wholesale electricity  market. } Furthermore, we show that forecast errors influence the volatility of intraday prices in a non-linear manner. 	
	
	{The paper is organized as follows. In the second part of the present section we comment on our idea and provide an intuitive description of our auction-curves-based model. The first part of section \ref{INPUTS} is devoted to the data description. The second part of section \ref{INPUTS} comments on a method to transform wholesale supply and demand curves into an equilibrium with a perfectly inelastic demand curve. Section \ref{Models} is dedicated to the description of our models. Section \ref{Resuls} presents the obtained results. More specifically, we discuss the results and features of our model, show the out-of-sample evaluation of the models and construct a numerical example to demonstrate the impact of forecast errors on the volatility of intraday prices. Furthermore, we elaborate on economic and policy implications of our research. Section \ref{COCNL} concludes the paper.}

	 \subsection{Main idea } \label{SUBSINT}

\begin{figure}[h]
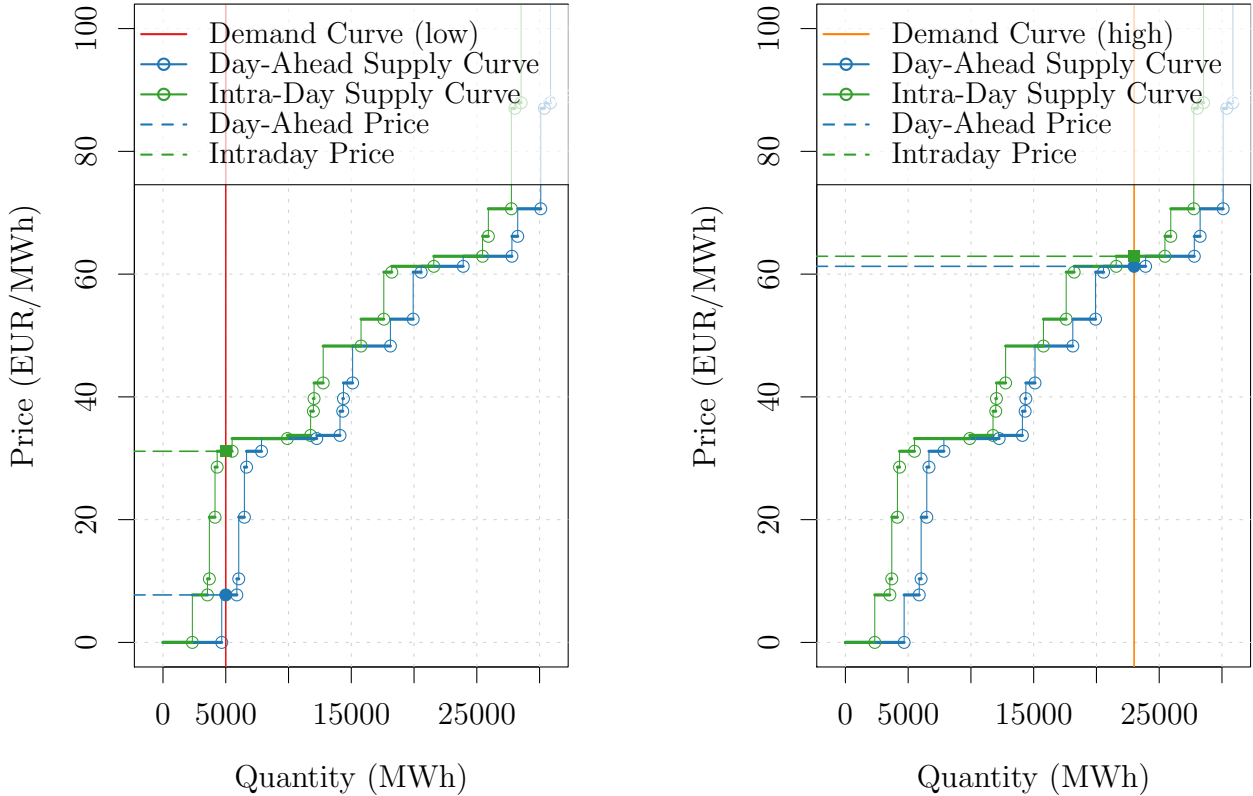

	\vspace{-1cm}
	\centering
	\subfigure[\normalsize Low demand scenario with demand ${D_{llow,t}=5000}$ MW \label{1FigureSub1}]
			{\scalebox{1}{\input{1FigureSub1.tex}}} \hspace{0.3cm}
	\subfigure[\normalsize High demand scenario with demand ${D_{high,t}=23000}$ MW \label{1FigureSub2}]
			{\scalebox{1}{\input{1FigureSub2.tex}}}
	%\resizebox{.99\textwidth}{!}{\input{1Figure.tex}}
	%	\vspace{-1.2cm}
	\caption{A toy example of an electricity market with the distance between day-ahead and intraday supply curves being dependent only on a negative forecast error of 2500 MW.}
	\label{FIG1}
\end{figure}

To elaborate on the main intuition behind our idea, we assume an imaginary toy example of an electricity market. This example is depicted in Figure \ref{FIG1}. We suppose that the blue curves denote supply curves recorded in a wholesale day-ahead market. The green curves are the corresponding hypothetical intraday supply curves. For the matter of simplicity we assume that the distance between the blue and green curves depends only on a forecast error of 2500 MW. Note that the blue curves are located to the right of the green curves. It follows that the actual amount of electricity was overestimated in the day-ahead market. Naturally, the blue curves would be shifted towards the green ones and the day-ahead price would be closer to the intraday price be the forecast in the day-ahead market more precise.

Note that the two curves, their shapes and the distances between them are identical on both sides of Figure \ref{FIG1}. The only difference between the two sides of the Figure is the realized demand size. The intraday price is more different to the day-ahead price in Figure \ref{1FigureSub1} (low demand case) than in Figure \ref{1FigureSub2} (high demand case) even though the size of the forecast error is the same. Figure \ref{FIG1} thus demonstrates that forecast errors may influence intraday prices in a non-linear manner.  As the Figure suggests, sector of a supply curve in which the prices are realized or the non-linear shape of the merit-order curve are factors which may determine the impact of forecast errors. %Recall now that that only forecast errors affect the distance between the curves. Hence, the day-ahead and the intraday prices would coincide if day-ahead and intraday errors would be equally large, i.e. if the two supply curves would overlay each other.

%As a benchmark we will use a naive model. This model tracks an unadjusted difference between intra-day and day-ahead prices. The more sophisticated linear regression model  considers the forecasting error $\varDelta$ and a couple of supplementary regressors. The third model is structurally similar to the latter one. It even has the same specification of the parameters. However, instead of using linear methods, it employs non-linear optimization techniques. We want to show that this model performs better than the previous one in order to confirm our initial idea. Finally, we combine linear and non-linear models with one another. We thus obtain the fourth model. We then compare the obtained models and observe that the latter one outperforms the other consistently. 

%More specifically, the price obtained by this model is an intersection between the demand function and an accordingly shifted theoretical supply curve, as analogous to the example given earlier.

%. In doing so, we shift the supply curve to the left or to the right depending on the forecasting error and then calculate an intersection between the demand and the shifted supply curves.

{Figure \ref{FIG1} can also be used to discuss the functioning of our novel auction-curves-based econometric model. The model is also optimization-based and is not underpinned by the analysis of day-ahead and intraday prices time series. Instead, the model is built around manipulations with empirical wholesale supply and demand curves. More specifically, our model shifts wholesale day-ahead supply curves to approximate the corresponding intraday supply curves. The magnitudes and the directions of the shifts are determined by (a) errors in wind and solar power forecasts and (b) absolute amounts of wind and solar power generated at the moment of delivery.\footnote{{We do not have the data regarding the wind and solar output at a moment shortly before the delivery. Thus, to carry out our study, we focus on the actually realized data. Hence, we can not use our model for intraday price forecasting. However, our model still allows us to trace the non-linear impact of forecast errors on intraday prices. }} To optimize the shift size, a non-linear optimization technique is applied. In other words, we add or subtract adjusted forecast errors from day-ahead supply. As a result, day-ahead auction curves are shifted horizontally. The shifted day-ahead supply curves are thus our approximations of intraday supply curves. Naturally, the intersections of the shifted day-ahead supply curves with the demand curves coincide with the intraday prices.\footnote{{The fact that intersections of the auction curves yield equilibrium prices is the core of fundamental  models (also known as structural models) elaborated in e.g. \cite{howison2009stochastic}.or \cite{carmona2013electricity}.} Moreover, \cite{ziel2016electricity} provided a lengthy analysis of the intersections of day-ahead wholesale auction curves. They showed that in 64 \% of the cases the intersections between the auction curves are identical to the reported prices, in 89 \% the error is less than 0.1 EUR and in 99.8 \% the error is less than 1 EUR. The reason for the errors to be present is e.g. block or other complex orders which are neglected by the model.} Therefore, our auction-curves-based model provides us with an innovative modeling approach of electricity prices. As opposed to conventional quadratic models, our model allows us to clearly interpret the impact of each of the model's parameters. Hence, the model is particularly interesting from the methodological perspective. Furthermore, our model still outperforms other models considered in the study.}

{In what follows we will compare our auction-curves-based model and its modifications with linear, quadratic and mixture benchmarks. We will use the same parameter specification in each of the models to ensure comparability. We will show that non-linear models can better approximate intraday prices. Moreover, we will demonstrate that our model outperforms the other models in the study during several hours of the day. The comparison of the models will thus allow us to conclude that the impact of forecast errors on intraday prices is non-linear. }

\section{Inputs of the Models} \label{INPUTS}
\subsection{Data description}
Following the introductory section of the present paper, we focus on the German-Austrian EPEX SPOT SE and study the period between 01.01.2016 and 31.12.2017 (see {\cite{epex2019curves} and \cite{epex2019daten}}). We will denote day-ahead prices by $P^{DA}_t$ and hourly weighted average intraday prices ({usually referred to as VWAP by the EPEX SPOT SE}) by $P^{ID}$.\footnote{Our choice of weighted average intraday prices is motivated by the findings of \cite{von2017optimal}. Following the conclusions of this paper, the majority of orders in the continuous intraday market arrive shortly before the gate closure.} As the regulation of the exchange suggests, prices in the day-ahead market are bound with $P_{\text{max}}=3000$ EUR  from above and with $P_{\text{min}}=-500$ EUR  from below. In turn, price range in the intraday market comprises -9.999 EUR to 9.999 EUR. Moreover, besides the price data, from the EPEX SPOT SE we also obtain the data regarding day-ahead wholesale supply and demand curves.  

Furthermore, from ENTSO-E Transparency ({see \cite{entsoe2019daten}}) we have the data regarding the forecasted and actual wind and solar power supply. We will index day-ahead forecasts of wind and solar generation by $F$ and the corresponding realized values by $A$. Hence, there are two pairs of parameters: $W^F$ and $W^A$, $S^F$ and $S^A$, where $W$ and $S$ stand for wind and solar power, respectively. {We compute forecast errors as $W^\varDelta = W^A- W^F$ and $S^\varDelta = S^A- S^F$}.

Please note that the data as to the renewable energy supply was provided in quarter-hourly resolution.{ We used simple arithmetic averages to adjust the granularity of the data to the hourly resolution}.  We did not extrapolate missing data and omitted the time points in which a price-volume observation was not available in at least one of the datasets. We neglected daylight saving times within the current research and did not make a clock-change adjustment.  Furthermore, we rounded the prices to two decimal places to spare computational time. 

\begin{figure}[h!]
	\centering
	\resizebox{.99\textwidth}{!}{\input{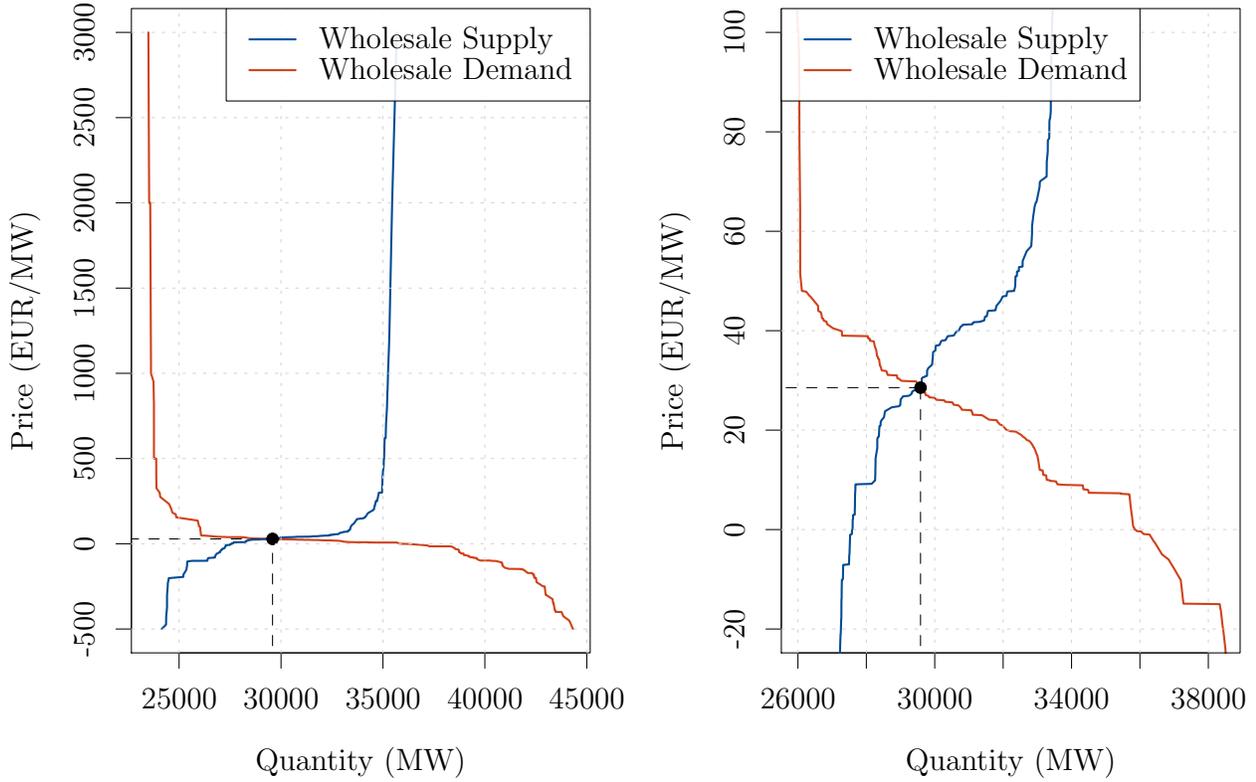}}
	\caption{Wholesale supply and demand curves in the German electricity market on 2017-04-02 at 09:00:00. The left hand side of the Figure shows the entire auction curves. The right hand side of the Figure shows the same two curves with a particular focus on their intersection. }
	\label{FIG2}
\end{figure}

\begin{figure}
	\input{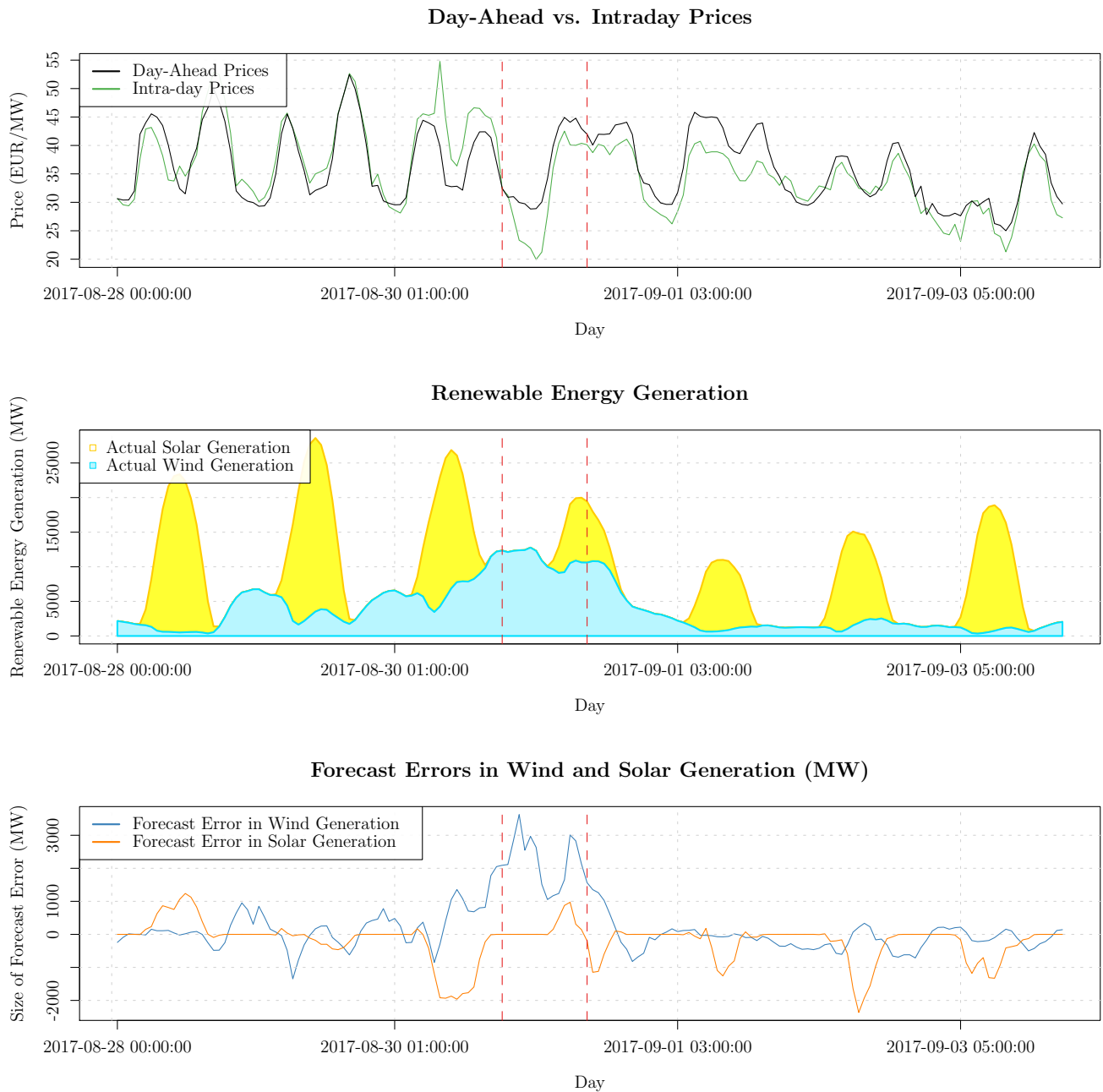}
	\vspace{-1cm}
	\caption{Dynamics of day-ahead and intra-day prices (upper plot), total generation of wind and solar energy (middle plot) and differences between actual and predicted wind and solar generation loads (lower plot) for a one-week sample from August, 28 to September, 03, 2017.}
	\label{3Graphs}
\end{figure}

Figures \ref{FIG2} and \ref{3Graphs} were constructed to present the data. The former Figure demonstrates an example of the wholesale supply and demand curves recorded in the German day-ahead market. The latter Figure shows the dynamics of day-ahead and hourly weighted average intraday prices, the total amount of wind and solar power supply and the forecast errors in wind and solar power output. From Figure \ref{3Graphs} it can be seen clearly that day-ahead prices may deviate from intra-day prices if the amount of wind or solar power was wrongly predicted. More specifically, the segment bounded by two red lines demonstrates explicitly that the intra-day prices can be smaller than the day-ahead prices (upper plot) when the actual amount of wind power (lower plot) was underestimated in day-ahead market.

		\subsection{Transformation of empirical supply and demand curves}

	%The method we use to construct our first non-linear model is analogous to that described in this paper and depicted in Figure \ref{FIG1}. Namely, we will subtract the amount $Z(\bbeta_{i:i+6})$ from the initial day-ahead supply forecasts. This will induce a day-ahead sale curve to be shifted to the left or to the right depending on the sign of $Z(\bbeta_{i:i+6})$. The magnitude of the shift will be additionally adjusted by solving a non-linear optimization problem and obtaining a vector of optimal $\bbeta_{i:i+6}$ coefficients. The shifted day-ahead sale curve thus acts as an approximation of an intraday sale curve. The intersection of the latter curve with the purchase curve coincides with a prognosis for an intraday price.  

	The toy example illustrated in Figure \ref{FIG2} plots a market equilibrium with the perfectly inelastic demand curve. Such setting allows us to shift the supply curves back and forward because market participants remain insensitive to the price. However, as Figure \ref{FIG1} demonstrates, the actual supply and demand curves are elastic. Shifting elastic curves may lead to ambiguous results because market participants may act differently depending on the equilibrium price. 
	
	To avoid this problem, we will use a method developed by \cite{coulon2014hourly}. This method allows us to transform the actual wholesale auction curves into a market equilibrium with perfectly inelastic demand. The equilibrium prices remain unchanged before and after the transformation, while the corresponding volume sizes increase. 
	 
	Economic reasoning behind the transformation was elaborated at length in the original paper or in e.g. \cite{kulakov2019determining}. Moreover, the paper by \cite{knaut2016hourly}, too, may ease the understanding of the underlying intuition. Speaking generally, the idea is to transfer all elasticities from the demand to the supply side. Due to the fact that the wholesale market offers opportunities for speculation, many market participants are trying to buy electricity in the wholesale market instead of generating it. As a result, many orders in the wholesale demand curve are of speculative nature. Therefore, even though being based on a somewhat relaxed assumption, the transfer of all demand elasticities to the supply side allows us to obtain a more stable and predictable market equilibrium with perfectly inelastic demand curve.
	 
\begin{figure}[h]
		\vspace{-1.5cm}
		\centering
		\subfigure{\input{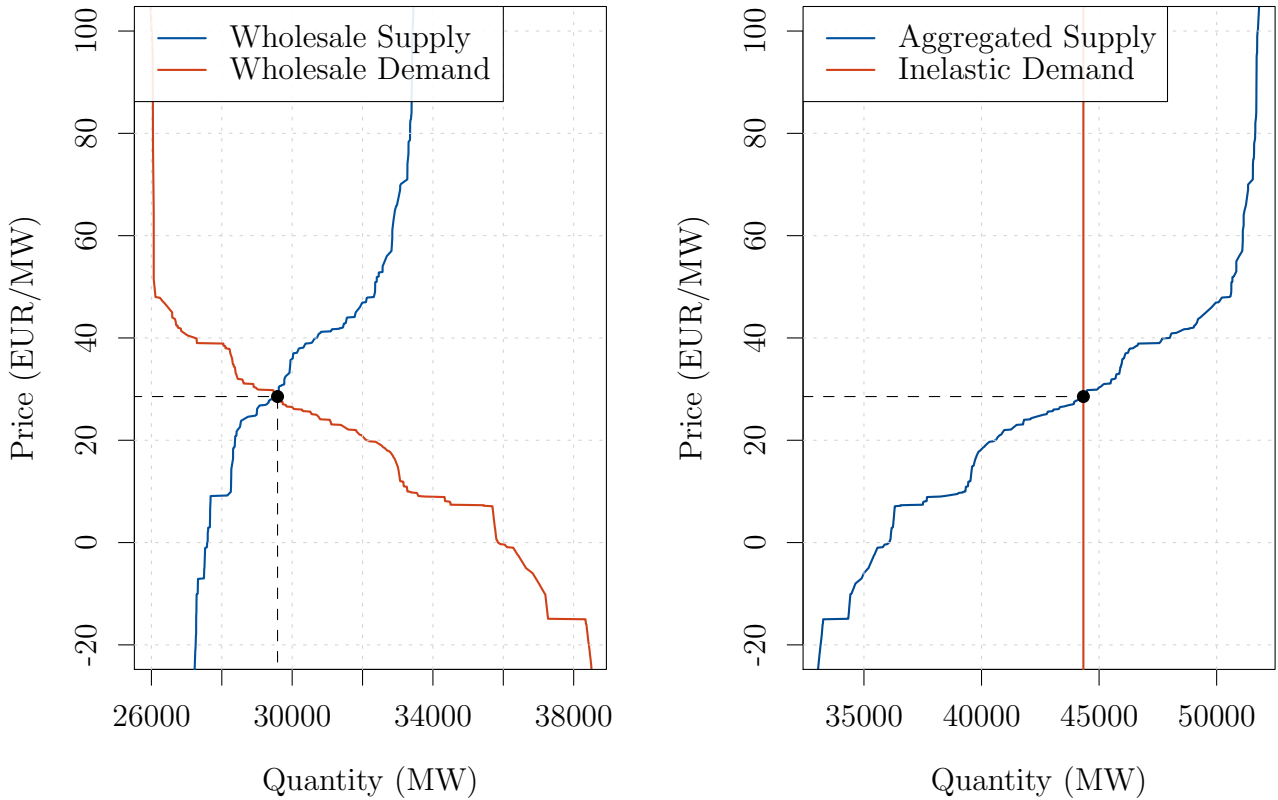}}		
		\caption{{A wholesale market equilibrium on 2017-04-02 08-00-00 CET (left plot) vs. its manipulated form with the inelastic demand curve (right plot).}}
		\label{FIG5}
	\end{figure}

	 Following the original paper, the expression for the inelastic demand curve can be written as 	
	\begin{equation}
	Dem^{inelastic}_t = WS Dem_t^{-1}(P_{\min})
	\end{equation}
	where a demand curve in a wholesale market is denoted by $WSDem$ and $P_{\min}=-500$ as follows from the regulation of the EPEX SPOT SE. In turn, the equation for an inverse supply curve can be written as
	\begin{align}
	Sup_t^{-1}(z) = WSSup_t^{-1}(z) + WSDem^{-1}_t(P_{\min}) - WSDem^{-1}_t(z)
	\end{align}
	where a supply curve in a wholesale market is denoted by $WSSup$. Please note that the above equation defines $Sup_t(z)$ automatically since the function is monotonic. Note that an example of the transformed wholesale auction curves is illustrated in Figure \ref{FIG5}.

	 % Hence, prior to modeling intraday prices by shifting corresponding forecasted day-ahead sale curves, we will first transform the initial sale and purchase curves according to the method mentioned above. 
	
	\section{Methodology} \label{Models}

	\subsection{Benchmark models}
	Benchmark models we use in the present study are state-of-the-art approaches in the field of intraday electricity price forecasting (see e.g.  %\cite{uniejewski2017variance}, \cite{maciejowska2016probabilistic}, \cite{uniejewski2016automated} or 
	\cite{narajewski2019econometric} or \cite{uniejewski2019understanding}). However, we neglect autoregressive components in our models because we only want to measure the impact of forecast errors on intraday prices. 
	
	The first benchmark model we introduce is a typical naive model $		P_t^{naive}=P^{DA}_t+\varepsilon_t,$
	where $P^{DA}$ stands for a day-ahead price and $\varepsilon_t$ is an error term. The other models in our paper are based on a component denoted by $\bZ_t$ which is written in a vector form and includes 6 elements
	\begin{align}
	\bZ_t & = \left(W^{\varDelta-}_t, W^{\varDelta}_t, S^{\varDelta-}_t, S^{\varDelta}_t, W^A_t,  S^A_t \right)'		
	\end{align}
	where $W^\varDelta_t$ and $S^\varDelta_t$ are errors in wind and solar supply forecasts, respectively; ${W^{\varDelta-}_t = \max(-W^\varDelta_t,0)}$ and $S^{\varDelta-}_t= \max(-S^\varDelta_t,0)$ stand for negative errors in wind and solar supply forecasts, respectively; $W^A_t$ and $S^A_t$ denote absolute volumes of the harvested wind and solar energy, respectively.  Note that we model negative forecast errors separately analogously to what was done in  e.g. \cite{soysal2017intraday}, \cite{kiesel2017econometric} or \cite{ziel2017modeling}. Vector $\bZ_t$ will thus be incorporated into several models and different estimation techniques will be used to determine the corresponding vectors of parameters $\bbeta=(\beta_0,...,\beta_i)'$.

	The first linear benchmark model $lm_1$ can be characterized as follows
	\begin{align}
		P_t^{lm_1} & = \beta_0 +\bbeta_{1:6}'\bZ_t + P_t^{DA}+ \varepsilon_t . \label{LM1}
		\end{align}
	  The second linear model $lm_2$ is analogous to the first one, save for the fact that term $P_t^{DA}$ is assigned with its own $\beta$ coefficient. 
	\begin{align}
		P_t^{lm_2} & =  \beta_0 + \bbeta_{1:6}'\bZ_t + \beta_{7} P_t^{DA} + \varepsilon_t . \label{LM2}
		\end{align}
	It follows that the model in equation \ref{LM1} is a special case of the model in equation \ref{LM2} with $\beta_7=1$. {Moreover, the former model assumes a perfect cointegration of intraday and day-ahead prices.} In fact the use of two similar models is justified because there is no clear academic consensus about which of them usually shows a better performance (see e.g. \cite{soysal2017intraday} or \cite{narajewski2019econometric}). 
	
	The last benchmark model is an extension of model $lm_2$ with quadratic terms. We will denote this model by $qlm$ and make the following statement
	\begin{align}
	P_t^{qlm} & =  \beta_0 + \bbeta_{1:6}'\bZ_t + \beta_{7} P_t^{DA} + \bbeta_{8:13}' \left(\bZ_t \circ \bZ_t \right) + \beta_{14} (P_t^{DA})^2 + \varepsilon_t,
	\end{align}
	where $\circ$ denotes the Hadamard or entrywise product.

	\subsection{Auction-curves-based models} \label{Snlm}
	
	Following the intuition elaborated in section \ref{SUBSINT}, our auction-curves-based models do not focus on day-ahead and intraday price time series. Instead, our models are based on manipulations with wholesale auction curves. From this perspective, our models follow a novel approach, but extend the family of econometric models developed in e.g. \cite{ziel2016electricity},  \cite{dillig2016impact}, \cite{shah2018forecasting} or \cite{kulakov2019determining}. Moreover, our models use the same parameter specification as the benchmark models. Therefore, it is possible to compare our models with the benchmarks. 
	
	We will denote the first auction-curves-based  model by $nlm$ and make the following statement 
	\begin{align}
	{P^{nlm}_t (\bbeta_{15:21}) = Sup_t(Dem^{inelastic}_t - \beta_{15}  - {\bbeta}_{16:21}'\bZ_t)) + \varepsilon_t}
	\end{align}
	where the price is namely an intersection of the shifted day-ahead supply curve with the inelastic demand curve.  Furthermore,	to estimate the optimal vector of coefficients ${\bbeta_{15:21}}$, we solve a non-linear least squares problem
	$
	\widehat{\bbeta}_{nlm} =  \argmin_{\bbeta \in \mathbb{R}^7 } \big(P^{ID}_t-P_t^{nlm}(\bbeta_{15:21})\big)^2
	$. In doing so, {we use the built-in \texttt{R} optimizer \texttt{optim} with default settings.} %We execute the optimization function at least 3 times to ensure that the obtained results are trustworthy.

	Our second model $cm$ incorporates linear model $lm_2$ and auction-curves-based model $nlm$.  The price equation of the model can be written as follows
\begin{align} %CHECK SUPPLY CHANGED FROM -1 to NLM 1
P^{cm}_{t} (\beta_{0:7, 15:22})   & =  \beta_0 + \underbrace{\bbeta_{1:6}'\bZ_t + \beta_7 P^{DA}_t}_{\text{linear component}} + \beta_{22} \underbrace{Sup_t\big({Dem^{inelastic}_t - \beta_{15}  - {\bbeta}_{16:21}'\bZ_t\big)}}_{\text{{non-linear component}}}  + \varepsilon_t
\end{align}
where the linear component coincides with the price produced by linear model $lm_2$ and the non-linear component is the price $P^{nlm}$. Writing the corresponding non-linear least squares problem yields
%	\begin{align}
$	\widehat{\bbeta}_{cm} =  \argmin_{\beta \in \mathbb{R}^{16} } \big(P^{ID}_t-P_t^{cm}(\beta_{0:7,15:22})\big)^2.$
	
	\subsection{Mixture models}
	{Following e.g. \cite{ziel2018day}, simply combining price models may yield further improvements of their performance.} Therefore,  additionally to the linear and auction-curves-based models, we also consider some of their equally weighted linear combinations. Please note that we report only on those models which showed superior performance compared to non-mixture models. First, model $mlq$ is a mixture of linear model $lm_2$ and quadratic model $qlm$ with $P_t^{mlq} = 0.5 P_t^{lm_2} + 0.5 P_t^{qlm}$. We use  model $lm_2$ and not $lm_1$ since the former one shows better performance according to MAE and RMSE tests. Second model $mnq$ has equation $P_t^{mnq} = 0.5 P_t^{qlm} + 0.5 P_t^{nlm}$, 
%
%	\end{align}
	%This concludes the description of our models. 
%	
third model $mcq$ can be represented as $
	P_t^{mcq}=0.5 P_t^{cm} + 0.5 P_t^{qlm}.
	$
	\section{Results} \label{Resuls}
	
	\subsection{Model analysis}
	 
	The obtained $\beta$-coefficients for the year 2017 are summarized in Table \ref{TAB1}. Please note that $\beta$-coefficients of linear models $lm_1$ and $lm_2$ and quadratic model $qlm$ are primarily negative. These findings are consistent with those of e.g. \cite{ziel2017modeling}, \cite{kiesel2017econometric}, \cite{clo2015merit}, \cite{ketterer2014impact} or \cite{gurtler2018effect}  where the signs of the coefficients are similar to the ones we obtained.  Hence, when the sizes of e.g. negative forecast errors grow, market participants expect a lower electricity supply. As a result, market prices rise and  $\beta$-coefficients are negative.
	
	Yet, the coefficients of the auction-curves-based models are primarily positive. The higher the magnitude of e.g. a positive forecast error is, the more will the merit-order be shifted to the right, and thus the lower the prices are (see e.g. \cite{neubarth2006impact}, \cite{cludius2014merit}, \cite{ketterer2014impact}, \cite{kiesel2017econometric}, \cite{fursch2012merit} or \cite{roldan2016merit}). From this perspective, the intuition behind the functioning of both linear and auction-curves-based models is the same. %Moreover, please note that the coefficient $\beta_7$ in the model $lm_2$ and the sum $\beta_7+\beta_{22}$ in the model $cm$ are significantly different from one as indicated by corresponding t-tests (-5.7395 and -2.6509, respectively). 
	
\begin{figure}[h]
		\centering
		\scalebox{0.95}{\input{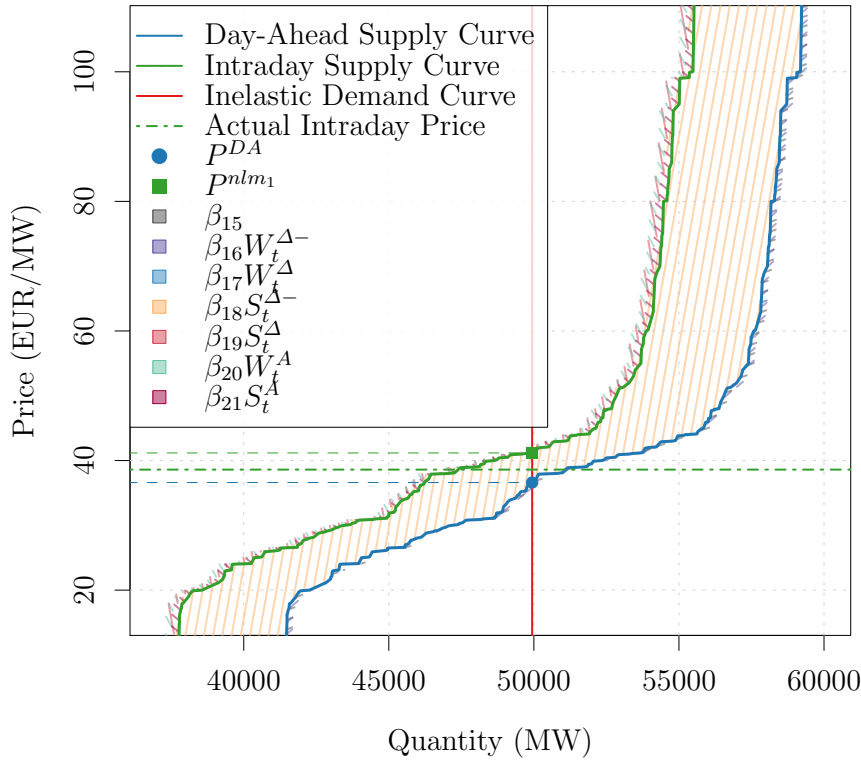}}		
		\caption{Price $P^{nlm}$ on 2017-04-19 at 10:00:00 as the result of shifting the transformed day-ahead supply curve to the left.}
		\label{FIG7}
	\end{figure}

		The application of model $nlm$ to the real data is illustrated in Figure \ref{FIG7}.  The areas highlighted in various colors demonstrate the shift sizes induced by the components of the model. The presence of the shaded areas to the left of the intraday supply curve indicates that the shift was partially negative. Furthermore, the intersection of the shifted supply curve with the inelastic demand curve yields the price $P^{nlm}$. 

\begin{figure}[h]
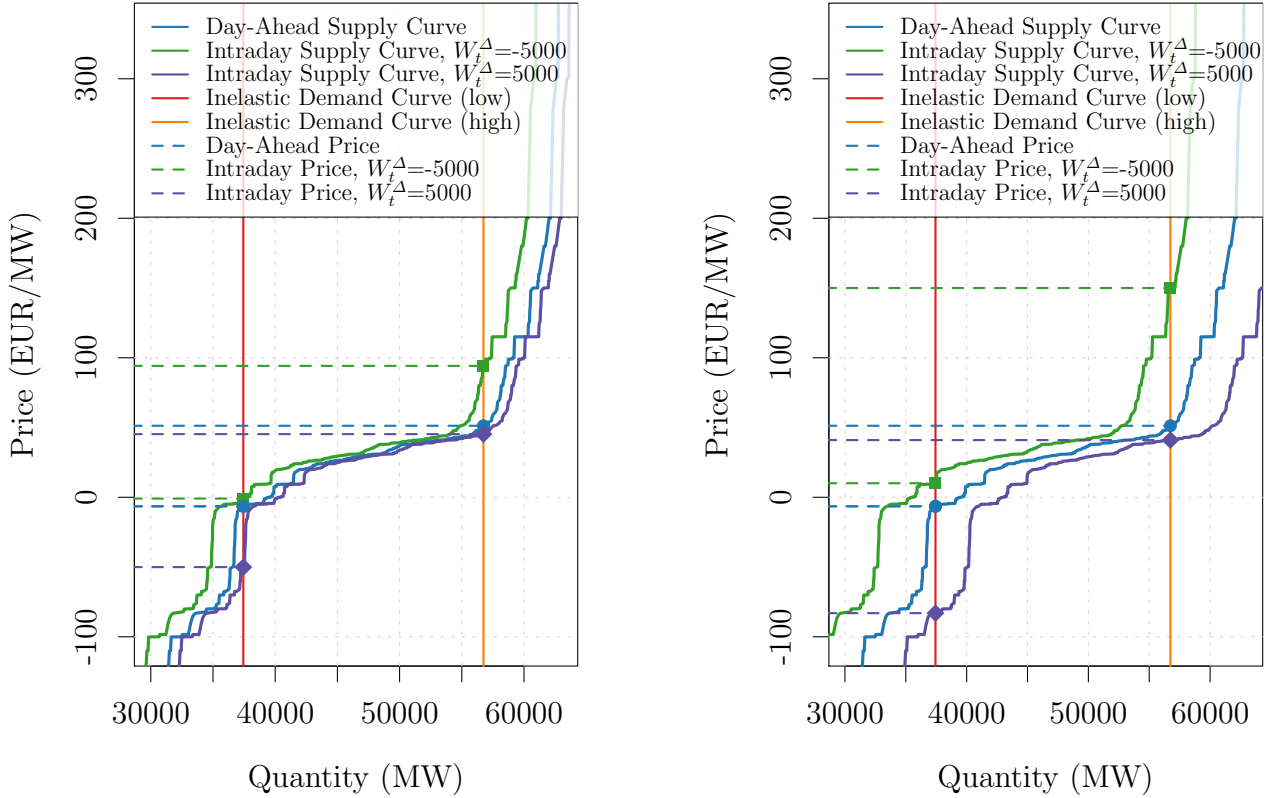

	\subfigure[\normalsize $W^\varDelta_t=-5000$ MW (green curve), $W^\varDelta_t=5000$ MW (purple curve) with $S^\varDelta_t=0$ MW  \label{LastGraphSub1} ]
	{\scalebox{1}{\input{LastGraphSup1.tex}}} \hspace{0.4cm}
	\subfigure[\normalsize $S^\varDelta_t=-5000$ MW (green curve), $S^\varDelta_t=5000$ MW (purple curve) with $W^\varDelta_t=0$ MW \label{LastGraphSub2}]
	{\scalebox{1}{\input{LastGraphSup2.tex}}}	
	\caption{{The day-ahead curve (blue) recorded on 2017-04-19 10:00:00 and its simulated shifts as results of positive and negative forecast errors for a low demand case with $D_{low,t}=37440$ MW and {$P_t^{DA}=-6.48$} and a high demand case with $D_{high,t}=56740$ MW and $P^{DA}_t=51.25$.}}
	\label{LastGraph}
\end{figure}

{Moreover, we constructed Figure \ref{LastGraph} as an example to show the differences in the impacts of positive and negative forecast errors on intraday prices. Both sides of the Figure plot the day-ahead supply curve (blue) recorded on 2017-04-19 at 10:00:00 and simulated shifts of this curve. The green curves show the shifts induced by negative forecast errors, the purple curves by positive. We used the framework of the model $nlm$ to build up the Figure. Both sides of the Figure incorporate a low demand scenario with $D_{low,t}=37440$ MW and a high demand scenario with $D_{high,t}=56740$ MW. We assume that the absolute amounts of wind and solar power output were equal to $W^A_t=15000$ MW and $S^A_t=15000$ MW. Figure \ref{LastGraphSub1} shows a scenario with forecast errors only in wind forecasts, i.e. $W^\varDelta_t=-5000$ MW, $W^\varDelta_t=5000$ MW and $S^\varDelta_t=0$ MW. On the contrary, Figure \ref{1FigureSub2} shows a scenario with $S^\varDelta_t=-5000$ MW, $S^\varDelta_t=5000$ MW and $W^\varDelta_t=0$ MW.}

{From Figure \ref{LastGraph} it appears clear that the horizontal distance between the green and blue curves is greater than the corresponding distance between the blue and purple curves. Following Table \ref{TAB1}, coefficients $W_t^{\varDelta-}$ and $S_t^{\varDelta-}$ are positive. Therefore, the impact of negative forecast errors on the shift of a day-ahead curve is greater than the impact of positive errors. As a result, the green curves   are moved further from the initial day-ahead supply curves. From the economic standpoint, we can argue that negative forecast errors can lead to the use of reserve capacities. Running additional power plants is  costly, which is why negative errors can exert a greater impact on intraday prices. }

{Moreover, it appears clear that horizontal distances between the green,blue and purple curves are greater in Figure \ref{LastGraphSub2} than in Figure \ref{LastGraphSub1}. In other words, an error in a solar power forecast causes a greater impact on intraday prices than an equally sized impact in a wind power forecast. However, large forecast errors (as e.g. of size $S^{\varDelta} = 5000$ MW or $S^{\varDelta} = -5000$ MW in our example) are relatively rare compared to similarly sized errors in wind forecasts due to a smaller amount of the generated solar output. {For example, in our sample (for both years 2016 and 2017) the absolute mean error (MAE) in wind forecasts was about 1000 MW, for solar 330 MW. Thus, the magnitude of an absolute average error in a solar forecast was roughly a third of that of a wind forecast. }} 

\begin{figure}[h]
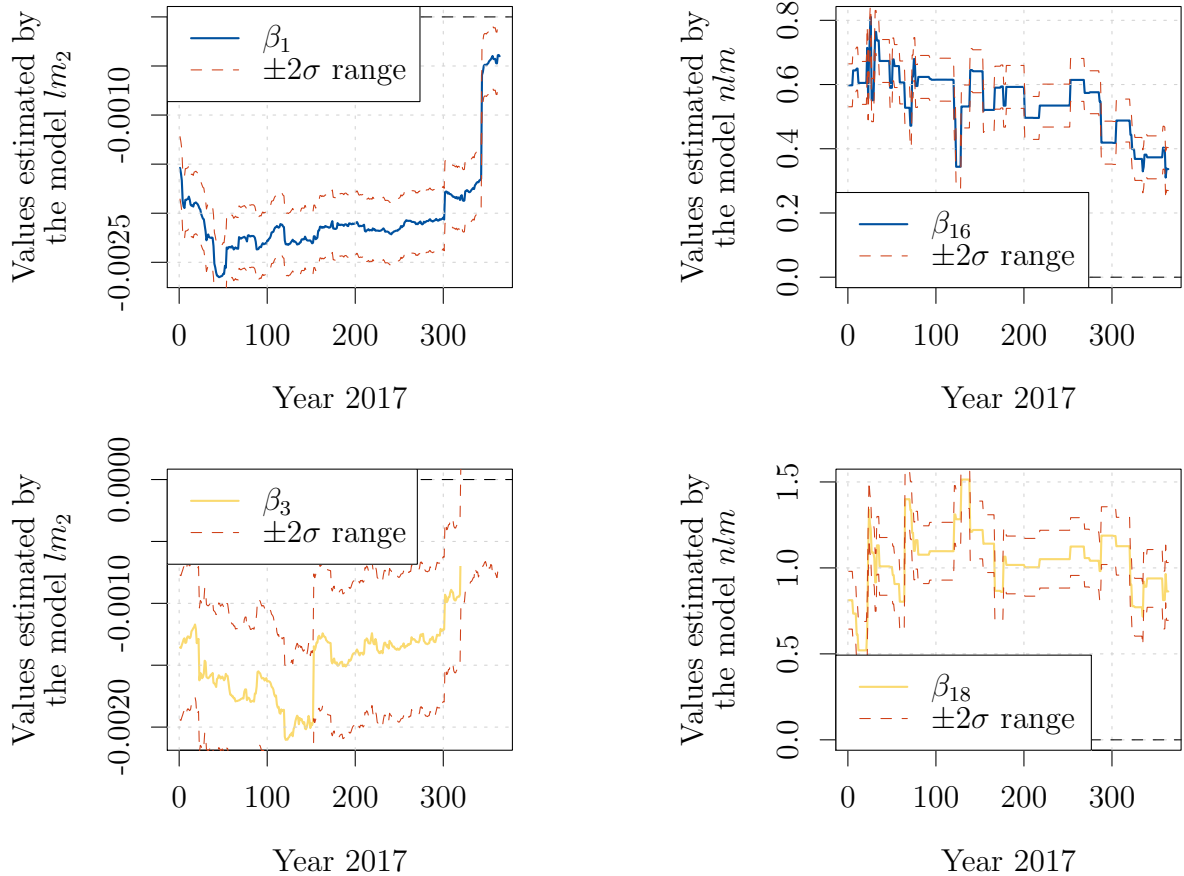

		\centering
		\vspace*{-1.5cm}
		\scalebox{1}{\input{WindAndSolarPrognoses1NoIAEE.tex}} 
		\begin{minipage}{\linewidth}
			\vspace{-1.5cm}
			\scalebox{1}{\input{WindAndSolarPrognoses2NoIAEE.tex}}
		\end{minipage}
		\caption{{$\beta$-coefficients for the negative parts of forecast errors for models $lm_2$ and $nlm$}}
		\label{FIG10}
	\end{figure}

	Furthermore, we also analyzed the asymmetries in the wind and solar power forecasts. Figure \ref{FIG10} plots the corresponding $\beta$-coefficients for the negative parts of the errors.  Please note that the coefficients of linear model $lm_2$ (left hand side of the Figure) are negative, the coefficients of auction-curves-based model $nlm$ (right hand side of the Figure) are positive. More importantly, both sides of Figure \ref{FIG10} show that the influence of negative forecast errors tends to drop over the year 2017. Similar findings are in e.g. \cite{gurtler2018effect}. 
	
	\subsection{Out-of-sample evaluation} 
	
	To evaluate the performance of the models, we first used the MAE and RMSE measures with the following specifications
	\begin{align}
	\text{MAE} = \frac{1}{24D}\sum_{d=1}^D \sum_{d=h}^{24} |P^{ID}_{d,h} - \what{P}^{ID}_{d,h}| \nonumber
	\intertext{and}
	\text{RMSE} = \sqrt{\frac{1}{24D}\sum_{d=1}^D \sum_{h=1}^{24}\left(P^{ID}_{d,h} - \what{P}^{ID}_{d,h}\right)^2}	\nonumber 
	\end{align}
	where $D=364$ and $h$ is a hour index. Hence, we used a rolling windows study with 365 in-sample observations (year 2016) and 364 out-of-sample observations (year 2017). The window size was equal to 24 hours. 	
	
	{The results of the MAE and RMSE tests are summarized in Table \ref{MAERMSE}. The Table allows us to see that model $lm_2$ has lower MAE and RMSE values than model $lm_1$. Model $lm_2$ was thus used in model $cm$. Furthermore, model $nlm$ fails to surpass the linear models and quadratic model $qlm$. Our model $cm$ performs better than model $qlm$ with respect to RMSE measure and has a very similar MAE value. Therefore, incorporating linear effects together with non-linear ones (as is done in model $cm$) allows us to obtain better results. In fact, combined model $cm$ shows best performance among the non-mixture models in our study. In turn, mixture model $mlq$ fails to beat model $qlm$ and the performance of mixture model $mnq$ and $mcq$ is best compared to the performance of the other models. } 
\begin{table}[h!]
	\fontsize{10pt}{12pt}\selectfont 
	\centering
	\begin{tabular}{l|l|l|l|l|l|l|l|l|l}
		\noalign{\hrule height 1pt}
		& Naive & $lm_1$ & $lm_2$ & $qlm$ & $nlm$ & $cm$ &$mlq$ & $mnq$& $mcq$  \\%& $qcm$ \\ 
		\noalign{\hrule height 1pt}
		MAE & 4.884 & 4.294 & 4.277 & {4.242} & 4.336 & 4.247& {4.244}  &4.198& \textbf{4.185} \\%\\ & 4.186 \\ 
		RMSE & 8.048 & 7.352 & 7.286 & 7.103 & 7.123 &  {7.097 }& 7.171 &\textbf{6.984}&7.010  \\ %\\ & 7.010 \\ 
		\noalign{\hrule height 1pt}
	\end{tabular}
	\caption{The results of the MAE and RMSE tests}
	\label{MAERMSE}
\end{table}

	To determine best model among the ones considered, we used the DM-test and applied it to each of the 24 hours separately.  We used DM-test formula from \cite{diebold2015comparing}. {To specify the parameters of the test, we defined the loss differential between models $\A$ and $\B$ for hour $h$ as $\delta_{d,h, \A,\B} = L_{d,h,\A} - L_{d,h,\B}$ where $L_{d,h}$ is the loss function of a model at hour $h$ of day $d$. {The respective loss functions of models $\A$ and $\B$ are $$L_{\A,d,h}=|\what{\varepsilon}_{\A,d+i,h}|^\varphi \hspace{0.5cm} \text{and} \hspace{0.5cm} L_{\B,d,h}=|\what{\varepsilon}_{\B,d+i,h}|^\varphi,$$} where $\varphi = 1,2$ to compare the models with respect to both absolute and quadratic errors. The t-statistics of the DM-test is given by $$
	t_{DM}=\frac{\bar{\delta}_{h,\A,\B}}{ \sigma_{\bar{\delta},h,\A,\B}}$$ % \sim \mathcal{N}_1(0,1) $$
	where $\bar{\delta}_{h,\A,\B}=\frac{1}{D}\sum_{d=1}^{D}\delta_{d,h,\A,\B}$ and $\sigma_{\bar{\delta},h,\A\B}$ is the standard deviation of $\bar{\delta}_{d,h, \A,\B}$. }
	{Figure \ref{FIG10} illustrates the hourly DM-test comparison of our best model $mcq$ against  best benchmark $qlm$. Given the 5\% confidence interval, we see explicitly that  model $mcq$ outperforms model $qlm$ during several hours of the day. Hence, we show that mixing our model with the quadratic benchmark allows us to outperform the other models in the study during several hours of the day.
		
	The above findings have the following implications. First, our model $mcq$ can be applied successfully to model intraday prices. Despite being unconventional,  the model yields similar (and even superior) performance relative to quadratic model $qlm$. Furthermore, the main advantage of  our model is the ease of interpretation. As opposed to quadratic model $qlm$, we can easily interpret the influence of each of the considered parameters by studying the contributions of each parameter to the shift size. Second, given the fact that the main components of our models were forecast errors in wind and solar power supply and absolute amounts of the wind and solar power generation, we can conclude that the impact of the forecast error on intraday prices is non-linear. This holds because model $qlm$ outperforms the linear ones, and because our mixture auction-curves-based model $mcq$ shows a better performance even compared to quadratic model $qlm$. {Hence, as was mentioned earlier, the non-linear shape of the merit order curve and a sector of this curve in which equilibrium price is realized are possible reasons for the non-linear impact of forecast errors on intraday electricity prices. In fact, the exact shape of the actual merit order curves remains unknown when both day-ahead and intraday prices are established. Therefore, higher forecast errors cause greater deviations of market prices from fundamental prices.}\footnote{\textcolor{black}{Following e.g. \cite{kyle1985continuous} or \cite{de2014market}, higher market liquidity implies lower risk premia. As a result, the impact of a higher forecast error on intraday prices may be lower in a deeper market. Yet, deriving a scientific proof of the statement is a subject of another study. }  } }

					\begin{figure}[h]
		\centering
		\vspace{-1cm}
		\scalebox{0.9}{\input{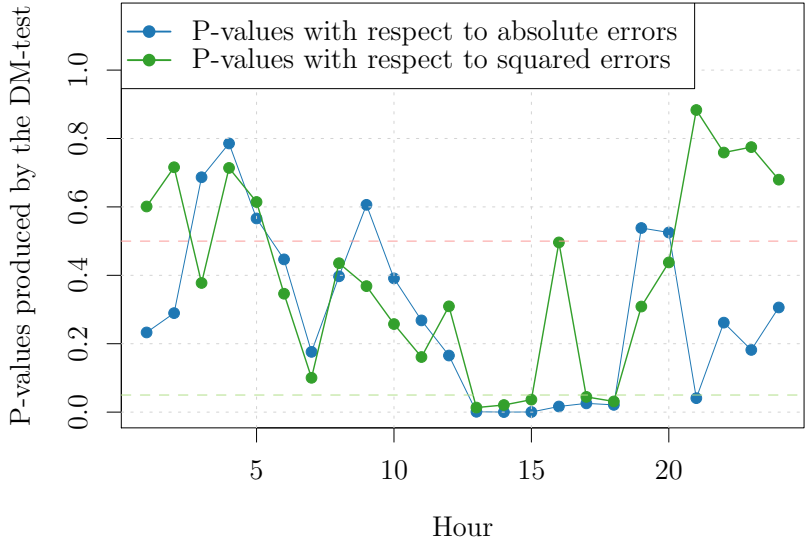}}
		\vspace{-0.4cm}		
		\caption{{Results of the DM-test comparison of models $mcq$ vs. $qlm$ for each hour of the day for the out-of-sample year 2017 }}
		\label{FIG9}
	
	\end{figure}

	\subsection{Forecast errors and volatility of intraday prices}
	
	%\cite{neubarth2006impact} %\cite{fursch2012merit} 

	Following e.g. \cite{clo2015merit}, additional wind and solar power capacities not only induced a merit-order effect in Italy, but also increased the volatility of electricity prices. These findings are consistent with those of \cite{woo2011impact} who conduct a similar study for the electricity market in Texas. {The authors of the paper show that the rising wind generation induces an increase in the variance of 15-minutes electricity spot prices.}  The corresponding analysis of the German electricity market is provided in \cite{ketterer2014impact}. The conclusions of the latter paper show that a rise in the volatility of electricity prices may stem from a growing penetration of renewable resources. 
	
%	Moreover, recall that it appears likely that the total amounts of wind and solar power plants will increase in the coming years. Therefore, the approach to building these new power plants must be adjusted accordingly for the negative impact of the errors not to be multiplied (see also e.g. \cite{weber2010adequate}). 

	%Note that the amount of wind and solar power capacities is expected to grow steadily in Germany (see e.g. \cite{henning2015will}). 
	In the present section we will develop a numerical example to show that the rising amount of wind or solar power capacities in fact increases the volatility of intraday prices. More importantly, our example demonstrates that the growth in the price volatility is driven by rising forecast errors and is non-linear. 
	
	Assume an operating onshore wind power plant and suppose that this power plant is extended with additional capacities.	Let $W_t$ be the amount of energy currently harvested by the plant, $\widetilde{W}_t$ an incremental wind supply from additional wind power capacities and $\gamma \geq 0$ a scale factor. As the conventional portfolio theory suggests (see e.g. \cite{berk2007corporate}), the generation variance of the extended wind power plant can  be computed as follows  
	\begin{align*}
	\Var[W_t+\gamma \widetilde{W}_t] & =% \Var[W_t] + 2\gamma \Cov\big[(W_t);(\widetilde{W}_t)\big] + \gamma^2 \Var[\wtilde{W}_t]\\
	\Var[W_t]  +  2\gamma \rho_{W_t,\widetilde{W}_t}\sqrt{\Var(W_t)\Var(\widetilde{W}_t)} +\gamma^2 \Var[\wtilde{W}_t]
	\end{align*}
	where $\rho$ denotes the correlation coefficient between the outputs of the old and new capacities. Assuming that variances $\Var[{W}_t]$ and $\Var[\wtilde{W}_t]$ are equal allows the above expression to be represented as follows
	\begin{align}
	%\Var[W_t+\gamma \widetilde{W}_t] & = \Var[W_t] \big( 1+ 2\gamma \rho_{[W_t;\widetilde{W}_t]} + \gamma^2 \big)  \Leftrightarrow \\
	\SD[W_t+\gamma\widetilde{W}_t] & = \SD[W_t]\underbrace{\sqrt{1+ 2\gamma \rho_{W_t,\widetilde{W}_t} + \gamma^2}}_{\text{greater than 1  if } \rho_{W_t,\widetilde{W}_t}>-\gamma/2} \label{SDWT}
	\end{align}
	Hence, the standard deviation of the electricity output of the combined power plant increases when $\rho>-\gamma/2$ (which is especially the case for $\rho>0$). Moreover, the unpredictability of volatile energy output implies that forecast errors,too, are high. In fact, \cite{weber2010adequate} suggests that the magnitude and amount of forecast errors rises inevitably together with expanding wind and solar power capacities.
	
	To show that  growing forecast errors may induce a non-linear increase in the volatility of electricity prices, we consider a numerical example.  As \cite{weber2010adequate} suggests, we can assume that forecast errors are proportional to the standard deviation of the power generation. Given this assumption, we can modify component {$\bZ_t$} to test the sensitivity of our models to changes in the amounts of wind and solar power capacities. In line with equation \ref{SDWT}, we suppose that our models are applied to a market with both old and additional wind and solar power capacities. We assume that modified component $\bZ_t$ is denoted by $\bZ_{t,\gamma}$ and can be represented as follows 
	\begin{align*}
	\bZ_{t,\gamma}& = \Big(W^{\varDelta-}_t  \sqrt{1+ 2\gamma_W \rho_{W_t,\widetilde{W}_t} + \gamma_W^2} \hspace{0.2cm}, \hspace{0.2cm} W^{\varDelta}_t   \sqrt{1+ 2\gamma_W \rho_{W_t,\widetilde{W}_t} + \gamma_W^2} \hspace{0.2cm},\\
	& S^{\varDelta-}_t   \sqrt{1+ 2\gamma_S \rho_{S_t,\widetilde{S}_t} + \gamma_S^2} \hspace{0.2cm},\hspace{0.2cm}  S^{\varDelta}_t   \sqrt{1+ 2\gamma_S \rho_{S_t,\widetilde{S}_t} + \gamma_S^2} \hspace{0.2cm}, 
	 W^A_t   \big(1 + \gamma_W\big) \hspace{0.2cm} ,\hspace{0.2cm} S^A_t   \big( 1+ \gamma_S \big) \Big) 
	\end{align*}
	To test our models under the new assumption, component $\bZ_t$ was replaced with $\bZ_{t,\gamma}$ in two of our models. We decided to compare our best linear model $lm_2$ with auction-curves-based model $nlm$ to demonstrate the differences between linear and non-linear settings. We have chosen $\gamma=0.1,1,5$ for the amounts of additional capacities and $\rho=0,0.5,0.8$ as possible correlation coefficients to conduct the study.  Table \ref{TAB2} summarizes the standard deviations of prices $P^{lm_2}$ and $P^{nlm}$ relative to the respective baseline models with $\gamma=0$ and $\rho=0$. Table \ref{TAB3} shows the differences between the true and the modeled intraday prices for the $0.1\%$ quantile relative to the respective baseline models with $\gamma=0$ and $\rho=0$. Table \ref{TAB4} shows the same as Table \ref{TAB3} but for the 99.9\% quantile. 
	
{	Note that the values in all three tables tend to increase the greater the additional power capacities are (i.e. the bigger $\gamma$ is) or the stronger the correlation between the old and the new capacities is (i.e. the higher $\rho$ is). More importantly, however, is that the values grow much quicker in model $nlm$. This indicates that the increase in the values is non-linear. For example, when considering Table \ref{TAB2}, non-linear effects be seen clearly by comparing the difference in the values produced by models $lm_2$ and $nlm$ for $\gamma=1, \rho=0.8$ and $\gamma=5, \rho=0.8$, especially for Wind+Solar case. Therefore, given the assumption that forecast errors are proportional to the standard deviations, Table \ref{TAB2} shows that the volatility of intraday prices increases when forecast errors grow. Hence, the numerical example allows us to conclude that forecast errors not only have a non-linear impact on intraday prices, but also influence the intraday price volatility in a non-linear manner. 
	
Moreover, Tables \ref{TAB3} and \ref{TAB4} offer another interesting conclusion. Table \ref{TAB3} shows that both in linear and non-linear models in the case of solar power the values can drop when $\gamma$ increases from $0.1$ to 1. This observation means that a moderate increase in the solar power capacities lowers the volatility of intraday prices at the 0.1\% quantile. The drop in the volatility happens because e.g. negative price spikes occur less often due to the increased solar output during the peak hours. A similar effect can be seen in Table \ref{TAB4} which focuses on the 99.9\% quantile. More specifically, the values in Table \ref{TAB4} can be lower than 1 in wind, solar and wind+solar cases. However, the observed effect is stronger in  linear model $lm_2$. It follows that moderate increase in wind and solar capacities can be beneficial for lowering the influence of positive price spikes.}

{Finally, the above example implies that the locations of new wind and solar power plants must be selected to minimize the correlations between the outputs of the old and new capacities. This issue has already been investigated on the grounds of empirical data in e.g. \cite{jonsson2010market} and \cite{grams2017balancing}. Moreover, our numerical example supports the findings of these papers. In fact, {there are several factors which may influence the strength of the correlation. For example, a new wind power plant can be built spatially close to the old one, or a new plant can be built in a place with similar weather conditions to the old one. As a result, fluctuations in wind output will influence the power supply of both old and new power plants simultaneously. Therefore, e.g. an unexpectedly small amount of wind in the system will almost equally affect the power supply of both old and new plants. Hence, the volatility of the power supply (and thus of electricity prices) increases. On the other hand, if the outputs of the old and new plants are negatively correlated, a drop in the output of one power plant will be offset by an increase of the output of another. The impact of extreme events on intraday prices and their volatility, too, is lower the lower the correlations between the old and new plants are. As a result, the overall volatility of electricity supply remains lower when the correlations are minimized. Hence, basic conclusions of conventional portfolio theory regarding portfolio diversification (see e.g. \cite{cochrane2009asset}}) hold in this context too. 
	}
	
	%The above results support our earlier claim regarding the policy and economical implications. Building new wind and solar power capacities and neglecting their correlations with old ones will not only boost the volatility of electricity generation. It will also lead to a greater amount or magnitude of forecast errors. This, in turn, will amplify the volatility of intraday prices in a non-linear manner. Hence, whenever the system stability is of importance, the .

	%A policy implication follows: in order to decrease the standard deviation of generation, new power plants should be built such that their generation has the lowest possible correlation with outputs of the already existing power plants. 

	\section{Conclusion} \label{COCNL} 
	
{	In this paper we studied the impact of errors in wind and solar power forecasts on  wholesale intraday electricity prices. To derive our conclusions, we elaborated a novel econometric model. Our model is based on manipulations with empirical supply and demand curves recorded in a wholesale electricity market. To compute the intraday price at a given time point, we horizontally shift the corresponding day-ahead supply curve. The magnitude and the direction of the shift depend on errors in wind and solar power forecasts and absolute amounts of wind and solar power. The shifted day-ahead supply curve is our approximation of intraday supply curve. The intersection of the approximated intraday supply curve with the demand curve coincides with the intraday price. Given that we can clearly see the contribution of each of the model's parameters to the shift size, the main advantage of our model is the ease of interpretation of the results. Moreover, given its novelty, our model can be particularly interesting from the methodological perspective. 
	
	More importantly, our results indicated that our auction-curves-based model outperforms other models in the study during several hours of the day. The quadratic benchmark, in turn, performs better than the linear benchmarks. \textcolor{black}{Hence, since the parameters in all of our models included only forecast errors and absolute amounts of wind and solar power, we could conclude that that the impact of forecast errors on intraday prices is non-linear.} Furthermore, on the grounds of a numerical example we showed that the impact of forecast errors on  the volatility of intraday prices is non-linear too. 
	
	Therefore, forecast errors may exert a strong impact on a market in case of a e.g. extreme event. Forecast errors shock not only intraday prices. The intraday price volatility, too, will react in a non-linear manner on forecast error. Of course, the increased volatility can be beneficial to some traders. However, from the perspective of a policy maker, higher uncertainty means higher risk premia and thus higher overall costs for an economy.  From this standpoint, to keep an energy market more stable (and thus to decrease the reliance on storage or reserve capacities), diversifying the supply of renewable energies can be an effective measure. Moreover, it can efficient to build additional wind and solar power capacities such that the correlations between the old and new plants is minimized. However, moderate increase in the wind and solar output may be beneficial for lowering the impact of price spikes. 
	
	Finally, further improvements of our auction-curves-based model are possible. For example, we only employed the shifts of the day-ahead supply curves to estimate the intraday supply curves. Additionally, we could also shift the demand curve. Moreover, using intraday data shortly before the delivery (and not the actually realized data) will allow the model to be used for intraday price forecasting.
	}

	 \newpage 
	 \section{Appendix}
	 
	 \textcolor{white}{space} 
	 	\begin{table}[h!]
	 	\fontsize{10pt}{12pt}\selectfont % Table font size
	 	\centering % Centered table
	 	\begin{tabular}{l|l|l|l|l|l|l}
	 		\noalign{\hrule height 1pt}
	 		& Multiplier & $lm_1$ & $lm_2$ & $qlm$ &$nlm$ & $cm$ \\ 
	 		\noalign{\hrule height 1pt}
	 		$\beta_{0}$ & $1$ & $-0.19777_{\star\star\star}$ & $\WM 1.24052^{***}$ & $\WM 2.46489^{***}_{\star\star\star}$ & $-$ & $\WM 0.10064$ \\ 
	 		
	 		$\beta_{1}$ & $W_t^{\varDelta-}$ & $-0.00039^*_{\star\star\star}$ & $-0.00040^{*}_{\star\star\star}$ & $\WM 0.00129^{***}_{\star\star\star}$ & $-$ & $\WM 0.00000_{\star\star\star}$ \\ 
	 		
	 		$\beta_{2}$ & $W_t^{\varDelta}$ & $-0.00214^{***}_{\star\star\star}$ & $-0.00209^{***}_{\star\star\star}$ & $-0.00410^{***}_{\star\star\star}$ & $-$ & $-0.00002_{\star\star\star}$ \\ 
	 		
	 		$\beta_{3}$ & $S_t^{\varDelta-}$ & $-0.00043_{\star\star\star}$ & $-0.00015_{\star\star\star}$ & $-0.00173^{*}_{\star\star\star}$ & $-$ & $\WM 0.00000_{\star\star\star}$ \\ 
	 		
	 		$\beta_{4}$ & $S_t^{\varDelta}$ & $-0.00258^{***}_{\star\star\star}$ & $-0.00273^{***}_{\star\star\star}$ &  $-0.00267^{***}_{\star\star\star}$ & $-$ & $-0.00002_{\star\star\star}$ \\ 
	 		
	 		$\beta_{5}$ & $W_t^A$ & $\WM 0.00009^{***}_{\star\star\star}$ & $\WM 0.00005^{***}_{\star\star\star}$ & $\WM 0.00014^{***}_{\star\star\star}$  & $-$ & $-0.00000_{\star\star\star}$ \\ 
	 		
	 		$\beta_{6}$ & $S_t^A$ & $\WM 0.00000_{\star\star\star}$ & $-0.00002_{\star\star\star}$ &  $\WM 0.00010^{*}_{\star\star\star}$ & $-$ & $-0.00000_{\star\star\star}$ \\

	 		$\beta_{7}$ & $P^{DA}_t$ & $-$ & $\WM 0.97019^{***}_{\star\star\star}$ & $\WM 0.86481^{***}_{\star\star\star}$ & $-$ & $\WM 0.39731^{*}_{\star\star\star}$ \\ 
	 		
	 		$\beta_{8}$ & $(W_t^{\varDelta-})^2$ & $-$ & $-$ & $\WM 0.00000^{***}_{\star\star\star}$ & $-$ & $-$ \\ 
	 		$\beta_{9}$ & $(W_t^{\varDelta})^2$ & $-$ & $-$ & $\WM 0.00000^{***}_{\star\star\star}$ & $-$ & $-$ \\ 
	 		$\beta_{10}$ & $(S_t^{\varDelta-})^2$ & $-$ & $-$ & $\WM 0.00000^{***}_{\star\star\star}$ & $-$ & $-$ \\ 
	 		$\beta_{11}$ & $(S_t^{\varDelta})^2$ & $-$ & $-$ & $\WM 0.00000_{\star\star\star}$ & $-$ & $-$ \\ 
	 		$\beta_{12}$ & $(W_t^A)^2$ & $-$ & $-$ & $\WM 0.00000^{***}_{\star\star\star}$ & $-$ & $-$ \\ 
	 		$\beta_{13}$ & $(S_t^A)^2$ & $-$ & $-$ & $\WM 0.00000^{*}_{\star\star\star}$ & $-$ & $-$ \\ 
	 		$\beta_{14}$ & $(P_t^{DA})^2$ & $-$ & $-$ & $\WM 0.00125^{***}_{\star\star\star}$ & $-$ & $-$ \\ 
	 		
	 		$\beta_{15}$ & $1$ & $-$ & $-$ & $-$ & $\WM 0.00004_{\star\star\star}$ & $-0.00061_{\star\star\star}$ \\ 
	 		
	 		$\beta_{16}$ & $W_t^{\varDelta-}$ & $-$ & $-$ & $-$ & $\WM 0.33663^{***}_{\star\star\star}$ & $\WM 0.90624_{\star\star\star}$ \\ 
	 		
	 		$\beta_{17}$ & $W_t^{\varDelta}$ & $-$ & $-$ & $-$ & $\WM 0.39478^{***}_{\star\star\star}$ & $\WM 0.24175^{***}_{\star\star\star}$ \\ 
	 		
	 		$\beta_{18}$ & $S_t^{\varDelta-}$ & $-$ & $-$ & $-$ & $\WM 0.86325^{***}_{\star\star\star}$ & $\WM 1.48092^{***}_{\star\star\star}$ \\ 
	 		
	 		$\beta_{19}$ & $S_t^{\varDelta}$ & $-$ & $-$ & $-$ & $\WM 0.37144^{***}_{\star\star\star}$ & $\WM 0.25544^{***}_{\star\star\star}$ \\ 
	 		
	 		$\beta_{20}$ & $W_t^A$ & $-$ & $-$ & $-$ & $-0.02659^{***}_{\star\star\star}$ & $-0.06149_{\star\star\star}$ \\ 
	 		
	 		$\beta_{21}$ & $S_t^A$ & $-$ & $-$ & $-$ & $-0.02590^{***}_{\star\star\star}$ & $-0.01337_{\star\star\star}$ \\

	 		$\beta_{22}$ & $P^{nlm}_t$ & $-$ & $-$ &$-$ & $-$ & $\WM 0.55152^{***}_{\star\star\star}$ \\ 
	 		\noalign{\hrule height 1pt}
	 	\end{tabular}
	 	\caption{The obtained $\beta$-coefficients, significance levels are: $\bullet$=10\% *=5\%, **=1\%, ***=0.1\% with respect to zero, $\circ$=10\%, $\star$=5\%, $\star\star$=1\%, $\star\star\star$=0.1\% with respect to one. } % Table caption
	 	\label{TAB1} % Table label for referencing the table elsewhere in the text with \ref{<label>}
	 \end{table}
 
 \newpage 
 
 	\begin{table}[ht]
 	\centering
 	\hspace*{0.5cm}
 	\begin{tabular}{l|l|l|l|l|l|l|l}
 		%	\cline{3-8}
 		\noalign{\hrule height 1pt}
 		\multicolumn{2}{c|}{}& \multicolumn{3}{c|}{Model $lm_2$} &\multicolumn{3}{c}{Model $nlm$}\\
 		\hline
 		$\gamma$ & $\rho$ & Wind & Solar & Wind+Solar & Wind & Solar & Wind+Solar \\ 
 		\noalign{\hrule height 1pt}
 		0.1 & 0.000 & 1.000 & 1.000 & 1.000 & 1.001 & 1.000 & 1.001 \\ 
 		0.1 & 0.500 & 1.000 & 1.000 & 1.000 & 1.001 & 1.001 & 1.001 \\ 
 		0.1 & 0.800 & 1.000 & 1.000 & 1.001 & 1.002 & 1.001 & 1.003 \\ 
 		1 & 0.000 & 1.016 & 1.004 & 1.020 & 1.026 & 1.006 & 1.029 \\ 
 		1 & 0.500 & 1.036 & 1.012 & 1.048 & 1.062 & 1.025 & 1.081 \\ 
 		1 & 0.800 & 1.051 & 1.018 & 1.069 & 1.089 & 1.034 & 1.124 \\ 
 		5 & 0.000 & 1.799 & 1.330 & 2.000 & 3.797 & 4.455 & 11.246 \\ 
 		5 & 0.500 & 1.931 & 1.397 & 2.161 & 4.245 & 5.032 & 13.274 \\ 
 		5 & 0.800 & 2.008 & 1.437 & 2.254 & 4.776 & 5.425 & 14.101 \\ 
 		\hline
 		0 & 0& \multicolumn{3}{c|}{4.980} &\multicolumn{3}{c}{4.974}\\
 		\noalign{\hrule height 1pt} 
 	\end{tabular}
 	\caption{Standard deviations of the prices $P^{lm_2}$ and $P^{nlm}$ relative to the respective baseline models with $\gamma=0$ and $\rho=0$.}
 	\label{TAB2}
 \end{table}
 
 \begin{table}[ht]
 	\centering
 	\hspace*{-0.2cm}
 	\begin{tabular}{l|l|l|l|l|l|l|l}
 		\noalign{\hrule height 1pt}
 		\multicolumn{2}{c|}{}& \multicolumn{3}{c|}{Model $lm_2$} &\multicolumn{3}{c}{Model $nlm$}\\
 		\hline
 		$\gamma$ & $\rho$ & Wind & Solar & Wind+Solar & Wind & Solar & Wind+Solar \\ 
 		\noalign{\hrule height 1pt}
 		0.1 & 0.000 & 1.005 & 1.000 & 1.005 & 1.047 & 1.000 & 1.047 \\ 
 		0.1 & 0.500 & 1.000 & 0.997 & 0.997 & 1.042 & 0.992 & 1.042 \\ 
 		0.1 & 0.800 & 0.997 & 0.996 & 0.993 & 1.036 & 0.986 & 1.036 \\ 
 		1 & 0.000 & 1.009 & 0.981 & 0.990 & 1.124 & 0.986 & 1.124 \\ 
 		1 & 0.500 & 0.977 & 0.963 & 0.940 & 1.059 & 0.986 & 1.068 \\ 
 		1 & 0.800 & 0.960 & 0.951 & 0.915 & 1.060 & 0.986 & 1.127 \\ 
 		5 & 0.000 & 1.077 & 1.043 & 1.232 & 1.851 & 2.166 & 3.158 \\ 
 		5 & 0.500 & 1.101 & 1.103 & 1.340 & 1.956 & 2.195 & 3.705 \\ 
 		5 & 0.800 & 1.129 & 1.136 & 1.401 & 2.438 & 2.200 & 4.009 \\ 
 		\hline
 		0 & 0& \multicolumn{3}{c|}{-39.190} &\multicolumn{3}{c}{-29.605}\\
 		\noalign{\hrule height 1pt} 
 	\end{tabular}
 	\caption{Differences between true and modeled intraday prices at the $0.1\%$ quantile relative to the respective baseline models with $\gamma=0$ and $\rho=0$.}
 	\label{TAB3}
 \end{table}

\begin{table}[ht]
	\centering
	\hspace*{-0.2cm}
	\begin{tabular}{l|l|l|l|l|l|l|l}
		\noalign{\hrule height 1pt}
		\multicolumn{2}{c|}{}& \multicolumn{3}{c|}{Model $lm_2$} &\multicolumn{3}{c}{Model $nlm$}\\
		\hline
		$\gamma$ & $\rho$ & Wind & Solar & Wind+Solar & Wind & Solar & Wind+Solar \\ 
		\noalign{\hrule height 1pt}
 	0.1 & 0.000 & 0.969 & 0.976 & 0.969 & 0.962 & 0.973 & 0.961 \\ 
 0.1 & 0.500 & 0.968 & 0.976 & 0.968 & 0.991 & 0.963 & 0.991 \\ 
 0.1 & 0.800 & 0.967 & 0.976 & 0.967 & 1.022 & 0.962 & 1.022 \\ 
 1 & 0.000 & 0.942 & 0.976 & 0.942 & 1.060 & 1.046 & 1.061 \\ 
 1 & 0.500 & 0.939 & 0.976 & 0.939 & 1.335 & 1.095 & 1.605 \\ 
 1 & 0.800 & 0.944 & 0.976 & 0.944 & 1.468 & 1.095 & 1.851 \\ 
 5 & 0.000 & 1.639 & 1.400 & 1.865 & 7.273 & 19.345 & 19.572 \\ 
 5 & 0.500 & 1.825 & 1.521 & 2.082 & 13.708 & 19.587 & 19.723 \\ 
 5 & 0.800 & 1.916 & 1.569 & 2.211 & 19.242 & 19.739 & 19.947 \\
		\hline
		0 & 0& \multicolumn{3}{c|}{25.504} &\multicolumn{3}{c}{26.135}\\
		\noalign{\hrule height 1pt} 
	\end{tabular}
	\caption{Differences between true and modeled intraday prices at the $99.9\%$ quantile relative to the respective baseline models with $\gamma=0$ and $\rho=0$.}
	\label{TAB4}
\end{table}

\newpage 

\newpage 
\bibliographystyle{apalike}
\bibliography{lib}
\end{document}